\documentclass[twocolumn,prb]{revtex4}
\usepackage{amsfonts}
\usepackage[T1]{fontenc}
\usepackage{amsmath,amsbsy,amssymb,graphicx}
\usepackage{times}

\begin{document}

\title{Electric circuits for universal quantum gates and quantum Fourier
transformation}
\author{Motohiko Ezawa}
\affiliation{Department of Applied Physics, University of Tokyo, Hongo 7-3-1, 113-8656,
Japan}

\begin{abstract}
Universal quantum computation may be realized based on quantum walk, by
formulating it as a scattering problem on a graph. In this paper, we
simulate quantum gates through electric circuits, following a recent report
that a one-dimensional $LC$ electric circuit can simulate a Schr\"{o}dinger
equation and hence a quantum walk. Especially, we propose a physical
realization of a set of universal quantum gates consisting of the CNOT,
Hadamard and $\pi /4$ phase-shift gates with the use of telegrapher wires
and mixing bridges. Furthermore, we construct the $\pi /2^{n}$ phase-shift
gate for an arbitrary integer $n$, which is an essential element to perform
the quantum Fourier transformation and prime factorization based on the Shor
algorithm. Our results will open a way to universal quantum computation
based on electric circuits.
\end{abstract}

\maketitle

Quantum computation\cite{Feynman,DiVi} is a most urgent and promising next
generation technique, which overcomes the limit of the Moore law. Universal
quantum computation is indispensable to make any quantum algorithms possible\cite{Deutsch}. 
According to the Solovay-Kitaev theorem, a set of the
universal gates consists of two Clifford and one non-Clifford gates\cite{Dawson}. 
The standard set is given by the controlled-NOT (CNOT), Hadamard
and $\pi /4$ phase-shift gates\cite{Universal}, where the $\pi /4$
phase-shift gate is a non-Clifford gate. Any unitary transformation is
executable with the use of these operators. One of the method to realize
universal quantum computation is based on quantum walks\cite%
{Child,Varba,Blumer,Hines,Lovett,Webb}, where widgets act as quantum gates.
However, it is still a long-standing problem how to achieve a physical
realization of these gates on the basis of quantum walks.

Recently, electric circuits attract much attention in the context of
topological physics\cite%
{ComPhys,TECNature,Garcia,Hel,Rosen,Lu,EzawaTEC,Hofmann,Research,EzawaLCR,EzawaMajo,HelSkin}. 
Majorana-like topological edge states can be simulated in electric circuits\cite{EzawaMajo}, 
and scalable topological quantum computation would be
possible based on their braiding\cite{EzawaTopCom}. On the other hand, it is
shown\cite{EzawaSch} that quantum walks are simulated by the telegrapher
circuit because the mathematical equivalence holds between the telegrapher
equation and the Schr\"{o}dinger equation.

\begin{figure}[t]
\centerline{\includegraphics[width=0.48\textwidth]{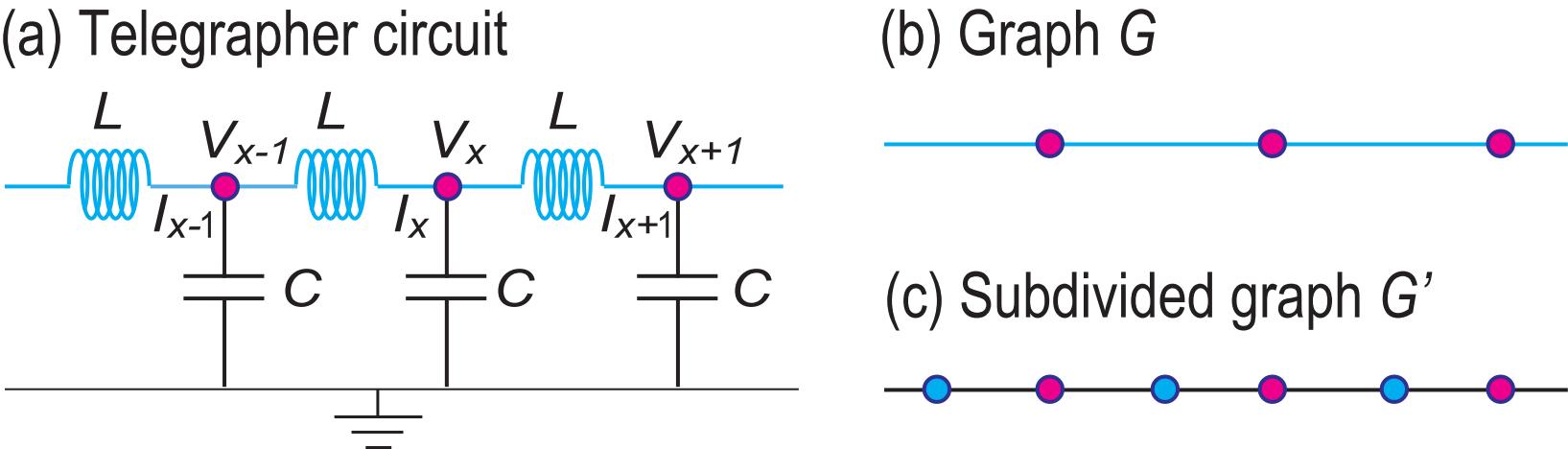}}
\caption{ (a) Illustration of a one-dimensional telegrapher circuit, where
two nodes (in magenta) are connected by inductor $L$ in cyan. Each node is
connected to the ground by capacitor $C$. (b) Illustration of the associated
graph $G$. (c) Illustration of the subdivided graph $G^{\prime }$, where
cyan vertices correspond to inductors $L$.}
\label{FigTeleWire}
\end{figure}

In this paper, we present a physical realization of a set of the CNOT,
Hadamard and $\pi /4$ phase-shift gates with the use of $LC$ electric
circuits. We consider a telegrapher circuit, which we map to a graph $G$ by
identifying nodes and links as vertices and edges, respectively. We
introduce a subdivided graph $G^{\prime }$ by transforming a link to a
vertex. The graph $G^{\prime }$ is bipartite since it involves two types of
vertices, representing voltage and current physically. The simplest example
is the one-dimensional telegrapher wire illustrated in Fig.\ref{FigTeleWire}(a), 
which corresponds to graphs $G$ and $G^{\prime }$ in Fig.\ref{FigTeleWire}(b) and (c), respectively. 
Quantum walks on graph $G^{\prime }$
have already been explored\cite{EzawaSch}. We note that, once a subdivided
graph is given, the corresponding electric circuit is readily designed.

\begin{figure}[t]
\centerline{\includegraphics[width=0.48\textwidth]{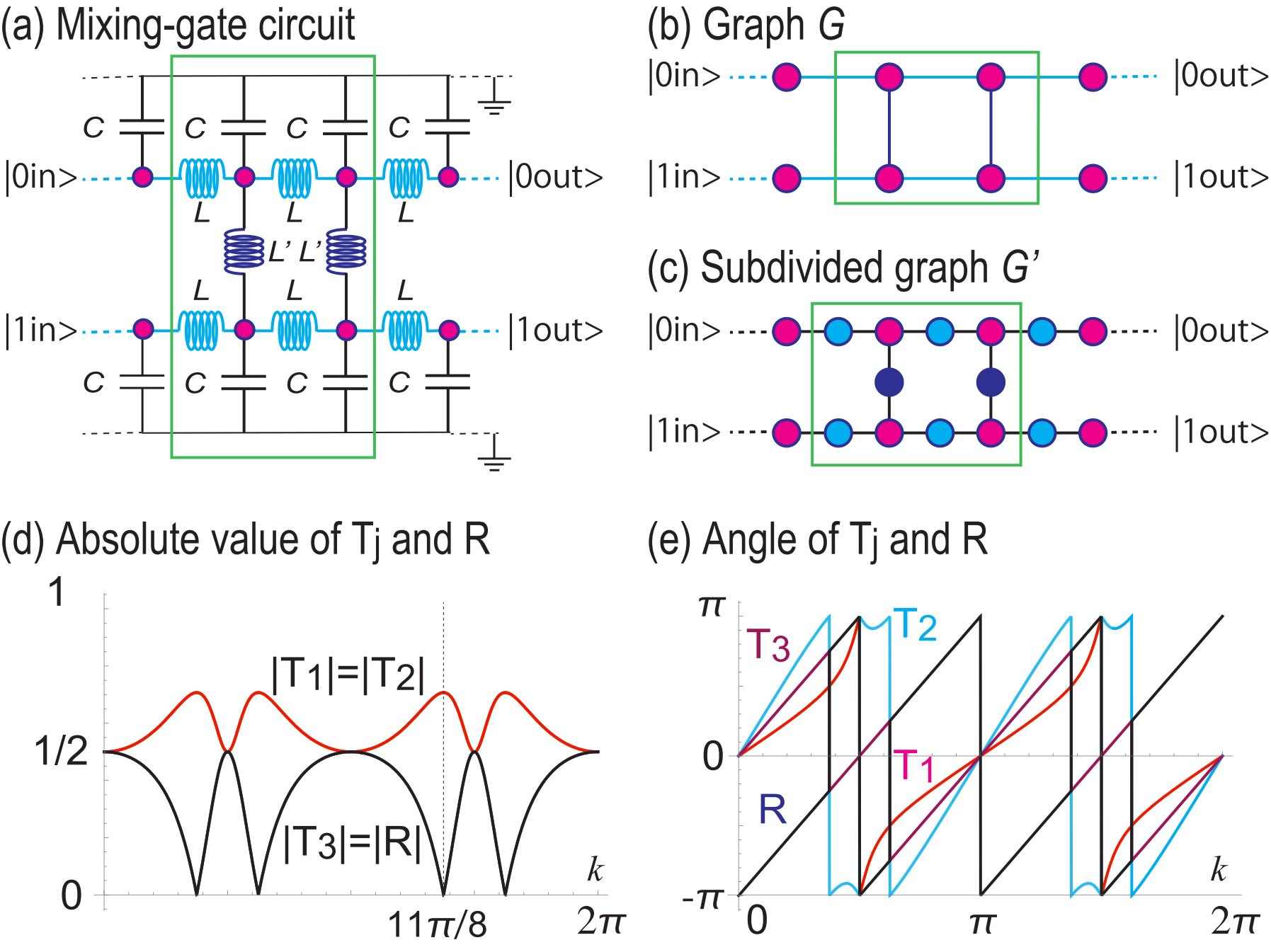}}
\caption{(a) Illustration of an electric circuit realizing the mixing gate $U_\text{mix}^{(2)}$ 
marked by a green rectangle. It consists of two parallel
wires bridged by two inductors $L^{\prime }$. (b) Graph $G$ for the mixing
gate. (c) The dynamics of the electric circuit is equivalent to a quantum
walk on the subdivided graph $G^{\prime }$. (d)--(e) The $k$ dependence of
the absolute value and the phase of the transmission coefficients $T_{1}(k)$, 
$T_{2}(k)$, $T_{3}(k)$ and the reflection coefficient $R(k)$. $T_{1}$ is
colored in red, $T_{2}$ is colored in cyan, $T_{3}$ is colored in violet and 
$R$ is colored in black. The horizontal axis is the momentum $0\leq k\leq 2\protect\pi $. 
We find $\left\vert T_{1}(k)\right\vert =\left\vert
T_{2}(k)\right\vert $ and $\left\vert T_{3}(k)\right\vert =\left\vert
R\left( k\right) \right\vert $.}
\label{FigMixing}
\end{figure}

\begin{figure}[t]
\centerline{\includegraphics[width=0.48\textwidth]{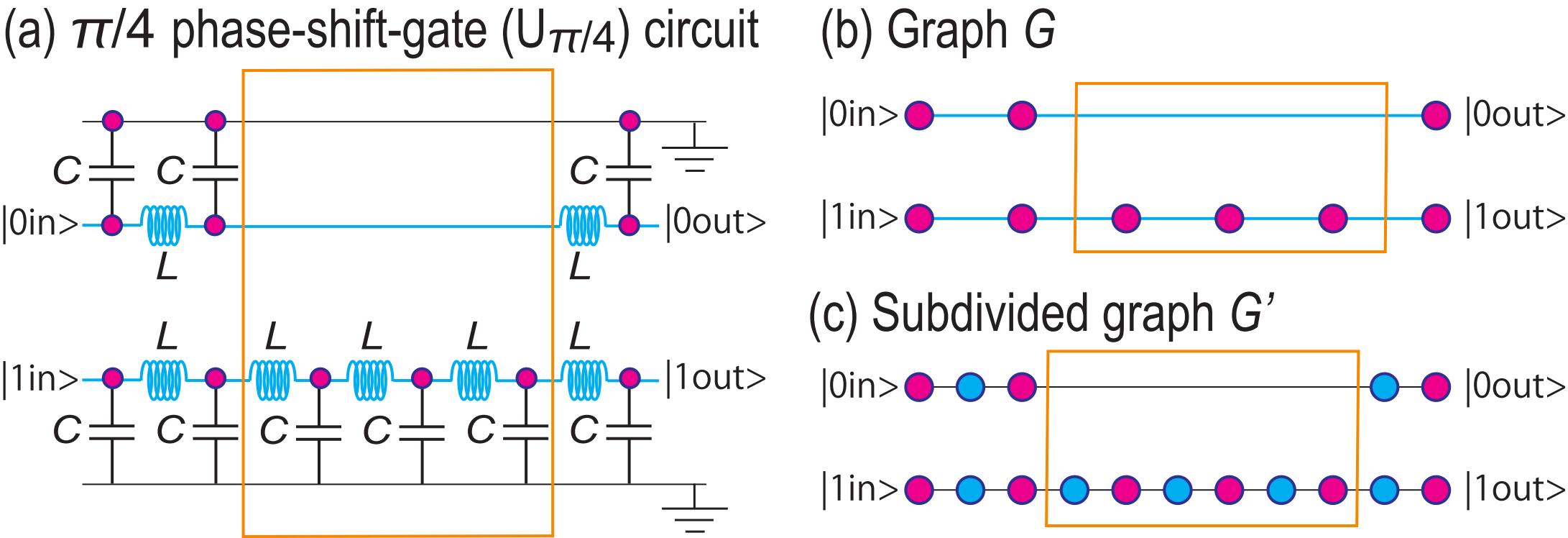}}
\caption{(a) Illustration of an electric circuit for the $\protect\pi /4$
phase-shift gate. Within the widget, there are no electronic parts in the
upper wire, while there are three inductors $L$ in the lower wire, producing
the phase delay $2mk_0$ with $m=3$ in the lower wire compared to the upper
wire. (b)--(c) Associated graphs. }
\label{FigPhaseShift}
\end{figure}

\begin{figure}[t]
\centerline{\includegraphics[width=0.48\textwidth]{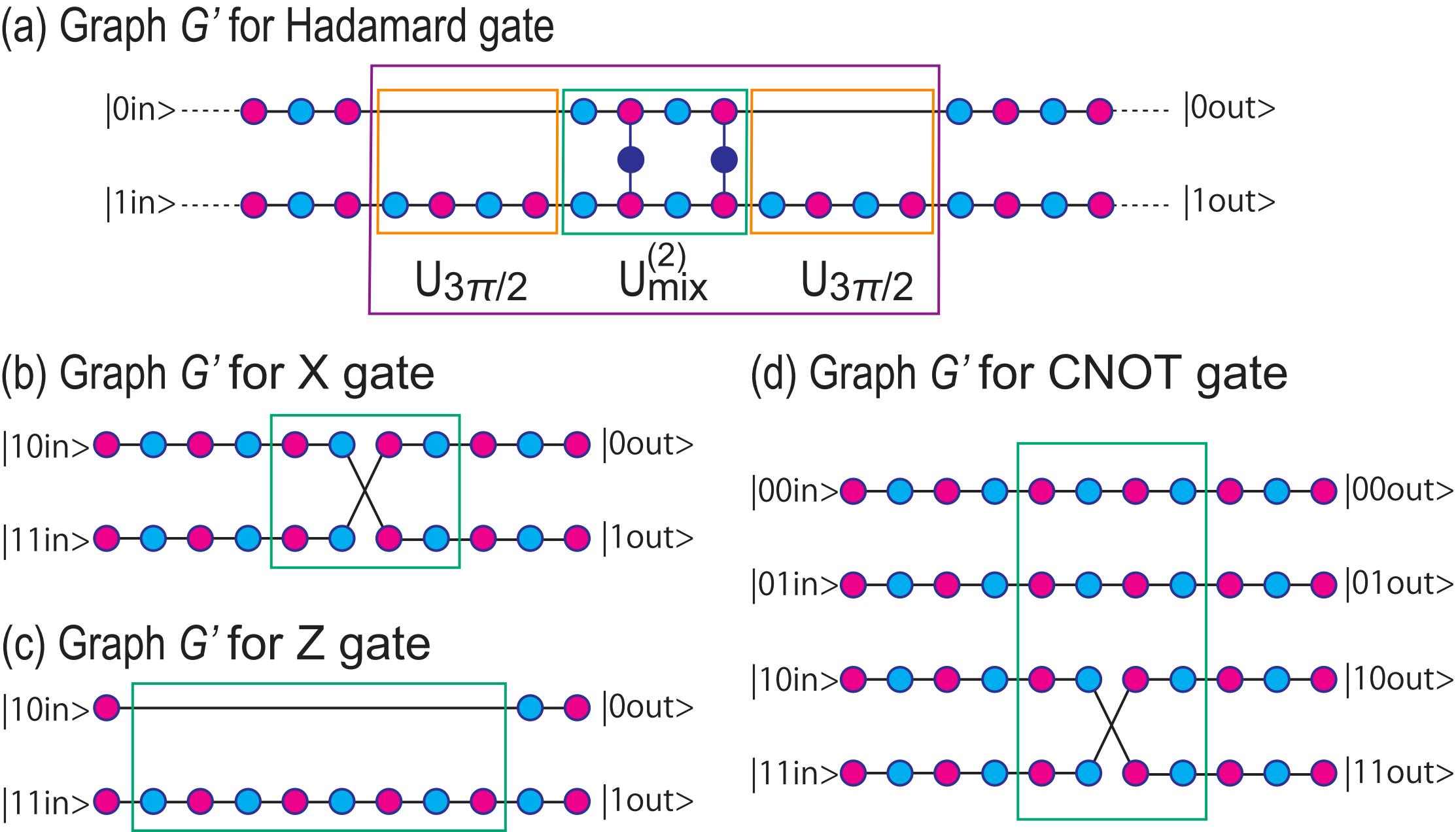}}
\caption{(a) Graph representation of the Hadamard gate, which consists of
the successive operations of the $3\protect\pi /2$ phase-shift and mixing
gates. (b)--(d) Graph representation of the Pauli X gate, the Pauli Z gate
and the CNOT gate.}
\label{FigHadaX}
\end{figure}

The next-simplest telegrapher circuit consists of four semi-infinite wires
for two inputs and two outputs, connected to a certain finite circuit called
a widget. It corresponds to a quantum gate represented by a $2\times 2$
matrix. There are three building blocks, the mixing gate $U_{\text{mix}}^{(2)}$, 
the $\phi $ phase-shift gate $U_{\phi }$ and the Pauli X gate $U_{X}\equiv \sigma _{x}$, where%
\begin{equation}
U_{\text{mix}}^{(2)}\equiv \frac{1}{\sqrt{2}}\left( 
\begin{array}{cc}
i & -1 \\ 
-1 & i
\end{array}
\right) ,\qquad U_{\phi }=\left( 
\begin{array}{cc}
1 & 0 \\ 
0 & e^{i\phi }
\end{array}
\right) .  \label{BasicU}
\end{equation}
The mixing gate corresponds to the one named "basis-changing gate" in
literature\cite{Child}. The phase-shift gate $U_{\phi }$ is constructed
simply by inserting inductors due to the induced electromotive force.
However, the mixing gate contains two bridges across two wires as in Fig.\ref{FigMixing}, 
and causes a back scattering of quantum walkers by the bridges.
Note that the superscript $(2)$ in $U_{\text{mix}}^{(2)}$\ implies the
number of the bridges. The suppression of back scattering is necessary for
quantum computing, which restricts the momentum associated with a quantum
walker. Possible values of the phase shift $\phi $ are determined by this
condition. The phase shift $\phi =\pi /4$ is such an allowed one, producing
the $\pi /4$ phase-shift gate as in Fig.\ref{FigPhaseShift}. The Hadamard
gate is constructed as in Fig.\ref{FigHadaX}(a) in accord with the relation 
$U_{\text{H}}=-iU_{3\pi /2}U_{\text{mix}}^{(2)}U_{3\pi /2}$, where $U_{3\pi
/2}=(U_{\pi /4})^{3}$. The X gate is simply given by exchanging two wires as
in Fig.\ref{FigHadaX}(b). The CNOT gate transforms four inputs to four
outputs, whose main part is the X gate as in Fig.\ref{FigHadaX}(d). The
standard set of univesal gates is constructed in this way.

\begin{figure}[t]
\centerline{\includegraphics[width=0.48\textwidth]{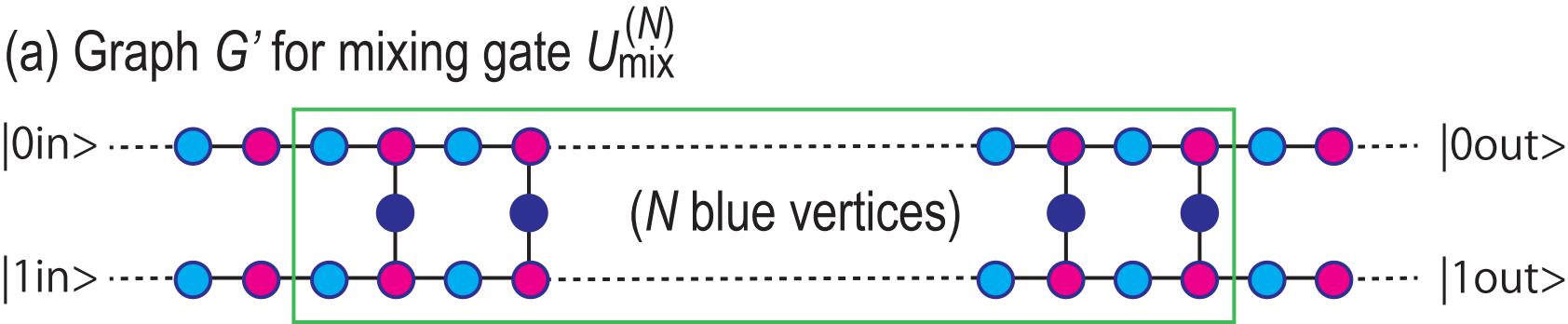}}
\caption{(a) Graph $G^{\prime }$ representation of the mixing gate $U_{\text{mix}}^{(N)}$ 
with $N=2^{n-1}$, which is composed of two wires bridged by $N$
inductors $L^{\prime }$ (in blue). It is used to construct the $\protect\pi/2^{n}$ phase-shift gate. }
\label{FigBridge}
\end{figure}

A most important application of quantum computers is prime factorization. A
key algorithm is the quantum Fourier transformation. It is decomposed into
the Hadamard gate $U_{\text{H}}$ and the $\pi /2^{n}$ phase-shift gate with
integer $n$ in (\ref{BasicU}). Although any quantum gate may be constructed
based on the standard set of universal gates, it is better to employ the 
$\pi /2^{n}$ phase-shift gate directly for practical purpose. We are able to
construct it together with the mixing gate $U_{\text{mix}}^{(N)}$ by
introducing $N$ bridges across two wires with $N=2^{n-1}$ as in Fig.\ref{FigBridge}.

Here we summarize the main differences between the previous studies\cite{Child,Varba,Blumer} 
and the present one as graph theory. First, the graph
is bipartite because it describes a network of voltage and current in the
electric-circuit realization. Second, the hopping parameters are pure
imaginary as in (\ref{HamilTB}),\ as originates in the mapping of the
telegrapher equation to the Schr\"{o}dinger equation. Third, the amplitude
of the hopping is not uniform because we use two types of inductors $L$ and 
$L^{\prime }$ as in Fig.\ref{FigMixing}, which makes the graph weighted.

\textit{Quantum walks and telegrapher equation:} Our basic idea is to use
telegrapher wires for a physical realization of quantum walks. A telegrapher
wire is a playground of one-dimensional quantum walkers\cite{EzawaSch},
where each node is connected to the ground by capacitor $C$, and to each
link an inductor $L$ is inserted, as illustrated in Fig.\ref{FigTeleWire}.
We also consider a widget connected to semi-infinite wires. For example,
there are two input wires and two output wires connected to a widget marked
in green in Fig.\ref{FigMixing}. The key problem is whether a quantum walker
may get through it without reflection.

An $LC$\ electric circuit is characterized by the Kirchhoff laws for the
voltage $V_{x}$ at the node $x$ and the current $I_{x}$ entering to the node $x$,%
\begin{align}
L_{x}\frac{d}{dt}I_{x}\left( t\right) =& V_{x^{\prime }}\left( t\right)
-V_{x}\left( t\right) ,  \label{EqB} \\
C_{x}\frac{d}{dt}V_{x}\left( t\right) =& I_{x}\left( t\right)
+\sum_{x^{\prime }}\left( -1\right) ^{s}I_{x^{\prime }}\left( t\right) ,
\label{EqA}
\end{align}%
where $x^{\prime }$ is one of the neighboring nodes of $x$ such that the
current is $I_{x}$ between these two nodes. The sign $s$ is determined
whether the current is in-going or out-going. The first equation is the
Kirchhoff voltage law with respect to the voltage difference between two
nodes $V_{x}$ and $V_{x^{\prime }}$, which originates in the induced
electromotive force by the inductor $L_{x}$. The second equation is the
Kirchhoff current law with respect to the current conservation at one node $V_{x}$. 
We denote a vertex representing the voltage $V_{x}$ (the current $I_{x}$) in magenta (cyan) in figures.

We may rewrite the Kirchhoff laws (\ref{EqB}) and (\ref{EqA}) in the form of
the Schr\"{o}dinger equation on the graph $G$,%
\begin{equation}
i\frac{d}{dt}\psi \left( x,t\right) =H\psi \left( x,t\right) ,  \label{Schro}
\end{equation}%
as we have presented a concrete example for one-dimensional quantum walks
elsewhere\cite{EzawaSch}. Here, the wave function is a two-component vector, 
$\psi \left( x,t\right) =\left( \mathcal{I}_{x}\left( t\right) ,\mathcal{V}_{x}\left( t\right) \right) ^{t}$, 
with $\mathcal{V}_{x}\left( t\right)=V_{x}\left( t\right) $ and $\mathcal{I}_{x}\left( t\right) =\sqrt{L/C}I_{x}\left( t\right) $. 
The Hamiltonian $H$ is given by the Kirchhoff laws
when an electric circuit is given.

We make separation of variables, $V_{x}\left( t\right) =\text{Re}[V\left(
x\right) e^{i\omega t}]$ and $I_{x}\left( t\right) =\text{Re}[I\left(
x\right) e^{i\omega t}]$, where $V\left( x\right) $ and $I\left( x\right) $
are complex voltage and current. Correspondingly, we define the complex
function $\psi \left( x\right) $ by $\psi \left( x,t\right) =\text{Re}[\psi
\left( x\right) e^{i\omega t}]$.

The aim of this paper is to propose a set of widgets which act as a set of
universal gates. The first step is to solve the Hamiltonian problem of an
infinite one-dimensional wire, where the Hamiltonian is given by\cite{EzawaSch} 
\begin{equation}
H=\frac{2}{\sqrt{LC}}\sum_{x}\left( i\left\vert \psi \left( x\right)
\right\rangle \left\langle \psi \left( x+1\right) \right\vert -i\left\vert
\psi \left( x+1\right) \right\rangle \left\langle \psi \left( x\right)
\right\vert \right) .  \label{HamilTB}
\end{equation}%
This Hamiltonian is diagonalized after Fourier transformation, 
$H=\sum_{k}E\left( k\right) \left\vert \psi _{k}\right\rangle \left\langle
\psi _{k}\right\vert $, where the eigen-energy is obtained as $E\left(
k\right) =(2/\sqrt{LC})\sin k$.

Then, we solve the Schr\"{o}dinger equation (\ref{Schro}) for a system where
semi-infinite wires are connected to a widget\cite{Child}. The key point is
to formulate the system as a scattering problem on a graph, where the wave
function is given by solving the eigen-equation\cite{Child,Varba,Blumer}%
\begin{equation}
H\psi _{k}\left( x\right) =(2/\sqrt{LC})\sin k\psi _{k}\left( x\right) ,
\label{EigenEq}
\end{equation}%
by requiring that the eigen-energy is the same as that of the semi-infinite
wire.

\textit{Scattering theory on graphs:} We start with a one-qubit gate $U$
from the input $(\left\vert 0\right\rangle _{\text{in}},\left\vert
1\right\rangle _{\text{in}})$ to the output $(\left\vert 0\right\rangle _{\text{out}},
\left\vert 1\right\rangle _{\text{out}})$. We realize it by a
set of two telegrapher wires containing a finite number of widgets, 
$U=U_{1}U_{2}\cdots $. We first analyze a widget illustrated in Fig.\ref{FigMixing}(a) explicitly, 
where two telegrapher wires are bridged by two
inductors $L^{\prime }$.

Linear electric circuits satisfy the superposition principle. Hence, it is
enough to calculate the transmission and reflection coefficients when we
input a plane wave only to the wire $\left\vert 0\right\rangle _{\text{in}}$. 
There are three other lines, where two of them are the outputs and the
rest is the other input. This is a scattering problem, and the wave
functions are written in the form of%
\begin{eqnarray}
\left\langle x,0|\psi \right\rangle &=&e^{-ikx}+R\left( k\right) e^{i\left(
k+\pi \right) x}, \\
\left\langle x,j|\psi \right\rangle &=&T_{j}\left( k\right) e^{ikx},
\end{eqnarray}%
where $T_{1}$ and $T_{2}$ are for the two outputs, while $T_{3}$ is for the
other input $j=1,2,3$. The momentum of the reflected wave is $k+\pi $ in
order to preserve the energy $E\left( k\right) $. In order to make a quantum
gate, it is necessary to determine the momentum $k=k_{0}$ by requiring 
$R\left( k_{0}\right) =0$ and $T_{3}\left( k_{0}\right) =0$ since the current
should not backflow to the two inputs.

By solving the eigen-equation (\ref{EigenEq}), we obtain%
\begin{eqnarray}
T_{1}\left( k\right) &=&\frac{e^{-2ik}+2\ell +e^{2ik}\left( 2\ell
^{2}-1\right) }{e^{-4ik}+e^{-2ik}4\ell +\left( 4\ell ^{2}-1\right) }, \\
T_{2}\left( k\right) &=&\frac{2\ell \left( e^{2ik}\ell +1\right) }{%
e^{-4ik}+4e^{-2ik}\ell +\left( 4\ell ^{2}-1\right) }, \\
T_{3}\left( k\right) &=&\frac{2\ell \left( \ell +\cos 2k\right) }{%
e^{-4ik}+4e^{-2ik}\ell +\left( 4\ell ^{2}-1\right) }, \\
R\left( k\right) &=&\frac{-2\ell \left( \ell +\cos 2k\right) }{%
e^{-4ik}+4e^{-2ik}\ell +\left( 4\ell ^{2}-1\right) },
\end{eqnarray}%
where $\ell =L^{\prime }/L$. The reflection becomes zero, $R\left(
k_{0}\right) =0$, at $k_{0}\equiv \pm \frac{1}{2}\arccos \left( -\ell
\right) +\eta \pi $ with $\eta =0,1$. At this momentum $k_{0}$, it follows
that $T_{3}\left( k_{0}\right) =0$, $T_{1}\left( k_{0}\right) =\pm i\sqrt{1-\ell ^{2}}$ 
and $T_{2}\left( k_{0}\right) =-\ell $. Hence we obtain%
\begin{equation}
U_{\text{mix}}^{(2)\pm }=\left( 
\begin{array}{cc}
\pm i\sqrt{1-\ell ^{2}} & -\ell \\ 
-\ell & \pm i\sqrt{1-\ell ^{2}}%
\end{array}%
\right)
\end{equation}%
for a possible widget acting as a quantum gate.

The Hadamard gate $U_{\text{H}}$ is given by%
\begin{equation}
U_{\text{H}}\equiv \frac{1}{\sqrt{2}}\left( 
\begin{array}{cc}
1 & 1 \\ 
1 & -1%
\end{array}%
\right) .  \label{Hadamard}
\end{equation}%
In order to construct it from $U_{\text{mix}}^{(2)\pm }$, we need to set 
$\left\vert T_{1}\left( k\right) \right\vert =\left\vert T_{2}\left( k\right)\right\vert $, 
which is achieved by choosing $\ell =1/\sqrt{2}$. Then, 
$U_{\text{mix}}^{(2)+}$ is realized at momenta $k_{0}=3\pi /8,11\pi /8$, 
while $U_{\text{mix}}^{(2)-}$ is realized at momenta $k_{0}=5\pi /8,13\pi /8$.

We show the transmission and reflection coefficients as a function of $k$\
for $\ell =1/\sqrt{2}$ in Fig.\ref{FigMixing}(d) and (e). We find 
$\left\vert T_{1}\left( k\right) \right\vert =\left\vert T_{2}\left( k\right)
\right\vert $ and $\left\vert R\left( k\right) \right\vert =\left\vert
T_{3}\left( k\right) \right\vert $ for all $k$.

\textit{Phase-shift gate:} We construct a phase-shift gate. It is simply
constructed by inserting $m$ inductors only in the lower wire compared to
the upper wire as shown in Fig.\ref{FigPhaseShift}. The transmission and
reflection coefficients are given by $T\left( k\right) =e^{2imk}$ and 
$R\left( k\right) =0$. The transmission is perfect irrespective of the
momentum since $\left\vert T\left( k\right) \right\vert =1$. Consequently,
we obtain the $\phi $ mixing-gate $U_{\phi }$ with $\phi =2mk$. It is
understood physically that inserted inductors cause delay in the lower wire.

Here the momentum for a phase-shift gate should be the same as that for the
mixing gate in one circuit. Then, we need to choose one momentum from 
$k_{0}=3\pi /8,11\pi /8,5\pi /8,13\pi /8$ for one circuit. Hereafter we
choose $k_{0}=11\pi /8$ since it requires the minimum number of inductors to
construct the $\pi /4$ phase-shift gate.

By using three inductors, i.e., $m=3$ and $k_{0}=11\pi /8$, we obtain the 
$\pi /4$ phase-shift gate as in (\ref{BasicU}). In general, by inserting $m$
inductors in one wire, we may generate the phase shift $\phi =2mk_{0}$. Two
successive operations of the $\pi /4$ phase-shift gate yield $U_{\pi
/2}=U_{\pi /4}^{2}=$diag.$(1,i)$, which is simply called the phase gate.

\textit{Hadmard gate:} The Hadamard gate is the unitary operation $U_{\text{H}}$ 
defined by (\ref{Hadamard}). It is given by the combination of the
mixing and $3\pi /2$ phase-shift gates as $U_{\text{H}}=-iU_{3\pi /2}U_{\text{mix}}U_{3\pi /2}$, 
where the $3\pi /2$ phase-shift gate is constructed
by inserting two inductors sequentially. We show the graph $G^{\prime }$ for
the Hadamard gate in Fig.\ref{FigHadaX}.

\textit{Pauli gates:} We study the NOT gate. It is defined by the
interchange of two inputs $\left\vert 0\right\rangle _{\text{in}}$ and 
$\left\vert 1\right\rangle _{\text{in}}$, or 
$\left\vert 1\right\rangle _{\text{out}}=U_{X}\left\vert 0\right\rangle _{\text{in}}$ and $\left\vert
0\right\rangle _{\text{out}}=U_{X}\left\vert 1\right\rangle _{\text{in}}$.
Such a gate is obviously given by the Pauli X gate, $U_{X}=\sigma _{x}$. In
terms of electric circuits, two telegrapher wires are simply interchanged as
shown in Fig.\ref{FigHadaX}(b).

The Pauli Z gate reads $U_{Z}=\sigma _{z}$. It is constructed by inserting
four successive inductors only for the lower wire as shown in Fig.\ref{FigHadaX}(c), 
because $2mk_{0}=\pi $ mod($2\pi $) with $m=4$. The Pauli Y
gate reads $U_{Y}=iU_{X}U_{Z}=\sigma _{y}$, which is constructed by the
successive operations of the Pauli Z and X gates.

\textit{CNOT gate:} The CNOT gate is a two-qubit gate $U$, transforming the
input $(\left\vert 00\right\rangle _{\text{in}},\left\vert 01\right\rangle _{%
\text{in}},\left\vert 10\right\rangle _{\text{in}},\left\vert
11\right\rangle _{\text{in}})$ to the output $(\left\vert 00\right\rangle _{%
\text{out}},\left\vert 01\right\rangle _{\text{out}},\left\vert
10\right\rangle _{\text{out}},\left\vert 11\right\rangle _{\text{out}})$. It
is defined by%
\begin{equation}
U_{\text{CNOT}}\equiv \left( 
\begin{array}{cc}
I_{2} & O_{2} \\ 
O_{2} & U_{X}%
\end{array}%
\right) ,
\end{equation}%
where $I_{2}$ is the two dimensional identity matrix and $O_{2}$ is the two
dimensional null matrix. Its construction is straightforward. We show the
graph for the CNOT gate in Fig.\ref{FigHadaX}(d).

$\pi /2^{n}$\textit{\ phase-shift gate:} We now construct the $\pi /2^{n}$
phase-shift gate for arbitrary integer $n$. First, we construct a mixing
gate which matches with the $\pi /2^{n}$ phase-shift gate, by generalizing
the result for the $\pi /4$ phase-shift gate. We bridge two telegrapher
wires by $N=2^{n-1}$ inductors with inductance $L^{\prime }$, as illustrated
in Fig.\ref{FigBridge}. The momentum $k_{0}$ and the inductance $L^{\prime
}=\ell L$ should be determined so that there are no reflections to the
inputs. The condition is satisfied when we choose $k_{0}=5\pi /4+\pi /(4N)$
and $\ell =-\cos 2k_{0}$, where the system acts as the quantum gate given by
\begin{equation}
U_{\text{mix}}^{(N)}=\frac{e^{i\pi N/2}}{\sqrt{2}}\left( 
\begin{array}{ll}
\exp \left[ -\frac{i\pi }{2}\left( \frac{1}{2}+\frac{1}{N}\right) \right] & 
\exp \left[ \frac{i\pi }{2}\left( \frac{1}{2}-\frac{1}{N}\right) \right] \\ 
\exp \left[ \frac{i\pi }{2}\left( \frac{1}{2}-\frac{1}{N}\right) \right] & 
\exp \left[ -\frac{i\pi }{2}\left( \frac{1}{2}+\frac{1}{N}\right) \right]%
\end{array}%
\right) ,
\end{equation}%
for $N\geq 4$. It is reduced to $U_{\text{mix}}^{(2)}$ in (\ref{BasicU}) for 
$n=2$ or $N=2$, where $k_{0}=11\pi /8$ and $\ell =-\cos 11\pi /4=1/\sqrt{2}$.

The $\pi /2^{n}$ phase-shift gate is constructed by inserting $m$ inductors
only in the lower wire as for the case of the $\pi /4$ phase shift gate,
yielding $U_{\phi }$ in (\ref{BasicU}) with $\phi =2mk_{0}$, where $m$ is
determined by the condition $2mk_{0}=\pi /2^{n}$ mod($2\pi $).

The Hadamard gate is constructed so as to realize the mathematical relation 
\begin{equation}
\exp \left[ \frac{i\pi }{2}\left( \frac{1}{2}+\frac{1}{N}+N\right) \right]
U_{3\pi /2}U_{\text{mix}}^{(N)}U_{3\pi /2}=U_{\text{H}},
\end{equation}%
where the $3\pi /2$ phase-shift gate\textbf{\ }$U_{3\pi /2}$\textbf{\ }is
constructed from\textbf{\ }the $\pi /2^{n}$ phase-shift gate as $U_{3\pi
/2}=(U_{\pi /2^{n}})^{3N}$.

In general we may construct the $\phi $ phase-shift gate $U_{\phi }$ for
arbitrary $\phi $\ from the $\pi /2^{n}$ phase-shift gate as $U_{\phi
}=(U_{\pi /2^{n}})^{s}$ with a certain integer $s$. It is well known that
the relation $s/N=2\phi /\pi $ holds within a required accuracy by taking
two integers $s$ and $N$ appropriately.

\textit{Quantum Fourier transformation:} The quantum Fourier transformation
is defined by\cite{Coppe}%
\begin{equation}
\left\vert k\right\rangle =\frac{1}{\sqrt{N}}\sum_{j=0}^{N-1}\omega
_{N}^{jk}\left\vert j\right\rangle
\end{equation}%
with $\omega _{N}=e^{2\pi i/N}$. It is constructed by successive operations
of the Hadamard and $2\pi /N$ phase-shift gates\cite{QBook}. Thus, we are
able to perform quantum Fourier transformation with arbitrary large $N$. See
some examples in Supplementary Material\cite{SM}.

Furthermore, it is possible to construct other multi-qubit gates such as the
CZ (controlled-Z), SWAP, Toffoli (controlled-controlled-NOT) and Fredkin
(controlled-SWAP) gates\cite{SM}. Various arithmetic operations can be made
by combinations of the NOT, CNOT and Toffoli gates\cite{Vedral}.

We have proposed a physical realization of quantum gates based on quantum
walks with the use of electric circuits. When we use inductors of the order
of 1nH and capacitors of the order of 1pF, the resonant frequency exceeds
1GHz. Then, we expect high speed quantum computations. Our results will open
a way to universal quantum computers made of relatively simple electric
circuits. Great merits are that we may control them by classical computers
seamlessly integrated and that they work at room temperature.

The author is very much grateful to E. Saito and N. Nagaosa for helpful
discussions on the subject. This work is supported by the Grants-in-Aid for
Scientific Research from MEXT KAKENHI (Grants No. JP17K05490, No. JP15H05854
and No. JP18H03676). This work is also supported by CREST, JST (JPMJCR16F1).

\clearpage\newpage
\onecolumngrid
\def\theequation{S\arabic{equation}}
\def\thefigure{S\arabic{figure}}
\def\thesubsection{S\arabic{subsection}}
\setcounter{figure}{0}
\setcounter{equation}{0}
%%%%%%%%%%%%%%%

\centerline{\textbf{\Large Supplemental Material}}
\bigskip
\bigskip

In the main text, we have constructed a set of the CNOT, Hadamard and $\pi
/4 $ phase-shift gates with the use of $LC$ circuits. Here we present graphs 
$G^{\prime }$ for some other typical gates, from which the corresponding $LC$
circuits are readily written down.

\textit{Square root of the NOT gate:} The square roof of the NOT gate $U_{X}$%
\ is given by%
\begin{equation}
U_{\sqrt{X}}\equiv \frac{1}{2}\left( 
\begin{array}{cc}
1+i & 1-i \\ 
1-i & 1+i%
\end{array}%
\right) ,
\end{equation}%
so that $U_{\sqrt{X}}^{2}=U_{X}$. It is realized by the sequential
application of the NOT gate and the mixing gates as%
\begin{equation}
U_{\sqrt{X}}=-e^{i\pi /4}U_{\text{mix}}^{(2)}U_{X}.
\end{equation}

\textit{One-qubit universal gate:} The one-qubit universal quantum gate is
constructed in the combination of the Hadamard and the phase-shift gates as 
\begin{equation}
U_{\text{1bit}}=e^{-i\theta /2}U_{\phi +\pi }U_{\text{H}}U_{\theta }U_{\text{%
H}}=\left( 
\begin{array}{cc}
\cos \frac{\theta }{2} & -i\sin \frac{\theta }{2} \\ 
ie^{i\phi }\sin \frac{\theta }{2} & -e^{i\phi }\cos \frac{\theta }{2}%
\end{array}%
\right) .
\end{equation}

\begin{figure}[b]
\centerline{\includegraphics[width=0.68\textwidth]{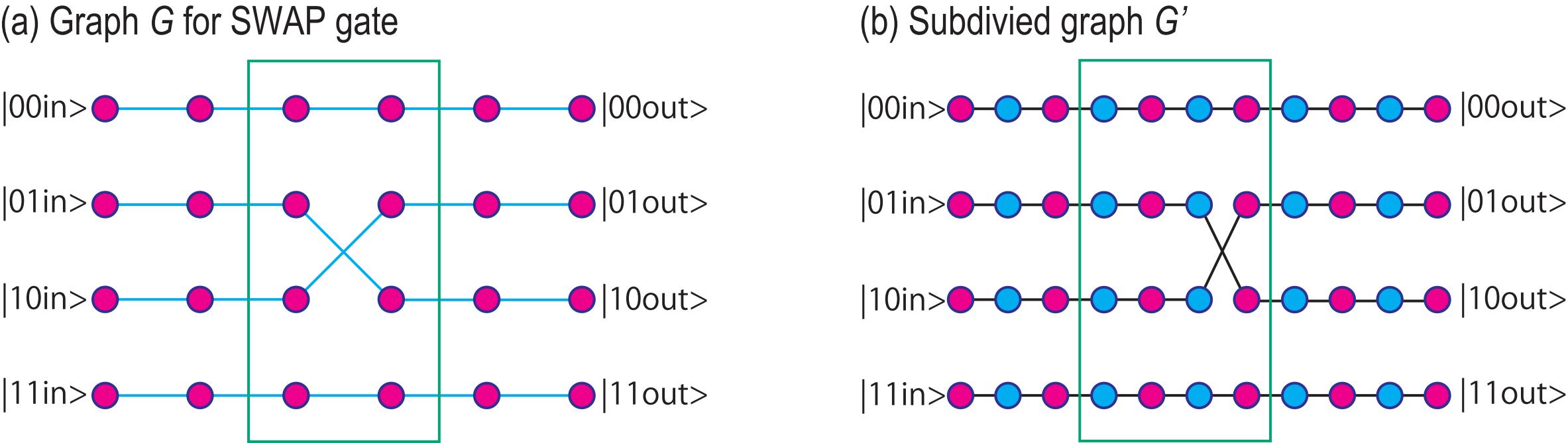}}
\caption{Graph representation of the SWAP gate, where the $\left\vert
01\right\rangle $ and $\left\vert 10\right\rangle $ are interchanged.}
\label{FigSWAP}
\end{figure}

\textit{SWAP gate:} The SWAP gate is defined by%
\begin{equation}
U_{\text{SWAP}}\equiv \left( 
\begin{array}{cccc}
1 & 0 & 0 & 0 \\ 
0 & 0 & 1 & 0 \\ 
0 & 1 & 0 & 0 \\ 
0 & 0 & 0 & 1%
\end{array}%
\right) .
\end{equation}%
It is constructed by the exchange of the wires between $\left\vert
01\right\rangle _{\text{in}}$ and $\left\vert 10\right\rangle _{\text{in}}$,
which are shown in Fig.\ref{FigSWAP}. The SWAP gate exchanges two qubits as 
\begin{equation}
U_{\text{SWAP}}\left\vert j_{2}j_{1}\right\rangle =\left\vert
j_{1}j_{2}\right\rangle .
\end{equation}

\begin{figure}[t]
\centerline{\includegraphics[width=0.68\textwidth]{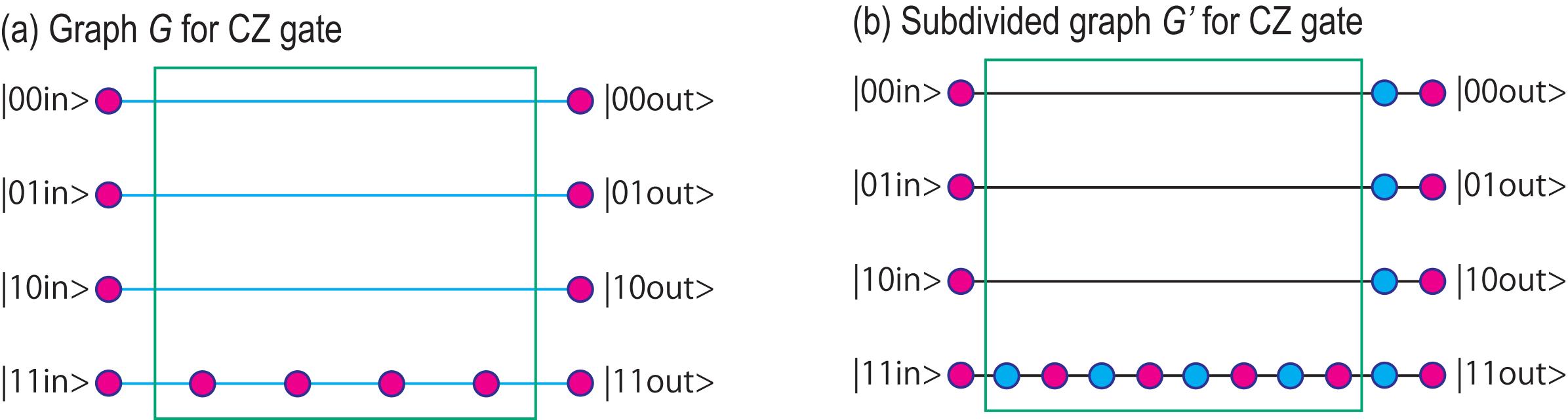}}
\caption{Graph representation on the CZ gate. It consists of the operation
of the $\protect\pi$ phase-shift gate only for $\left\vert 11\right\rangle$. 
}
\label{FigCZ}
\end{figure}

\textit{CZ gate:} The controlled-Z (CZ) gate is defined by

\begin{equation}
U_{\text{CZ}}\equiv \left( 
\begin{array}{cc}
I_{2} & O_{2} \\ 
O_{2} & U_{\text{Z}}%
\end{array}%
\right) =\left( 
\begin{array}{cccc}
1 & 0 & 0 & 0 \\ 
0 & 1 & 0 & 0 \\ 
0 & 0 & 1 & 0 \\ 
0 & 0 & 0 & -1%
\end{array}%
\right) .
\end{equation}%
We realize it by inserting four inductors only for the wire representing $%
\left\vert 11\right\rangle _{\text{out}}$.

\textit{Controlled phase-shift gate:} The controlled phase shift gate $%
U_{2\rightarrow \phi}$ is defined by 
\begin{equation}
U_{2\rightarrow \phi }\equiv \left( 
\begin{array}{cc}
I_{2} & O_{2} \\ 
O_{2} & U_{\phi }%
\end{array}%
\right) =\left( 
\begin{array}{cccc}
1 & 0 & 0 & 0 \\ 
0 & 1 & 0 & 0 \\ 
0 & 0 & 1 & 0 \\ 
0 & 0 & 0 & e^{i\phi }%
\end{array}%
\right) .
\end{equation}%
We realize it by inserting some inductors only to the wire representing $%
\left\vert 11\right\rangle _{\text{out}}$.

\begin{figure}[t]
\centerline{\includegraphics[width=0.68\textwidth]{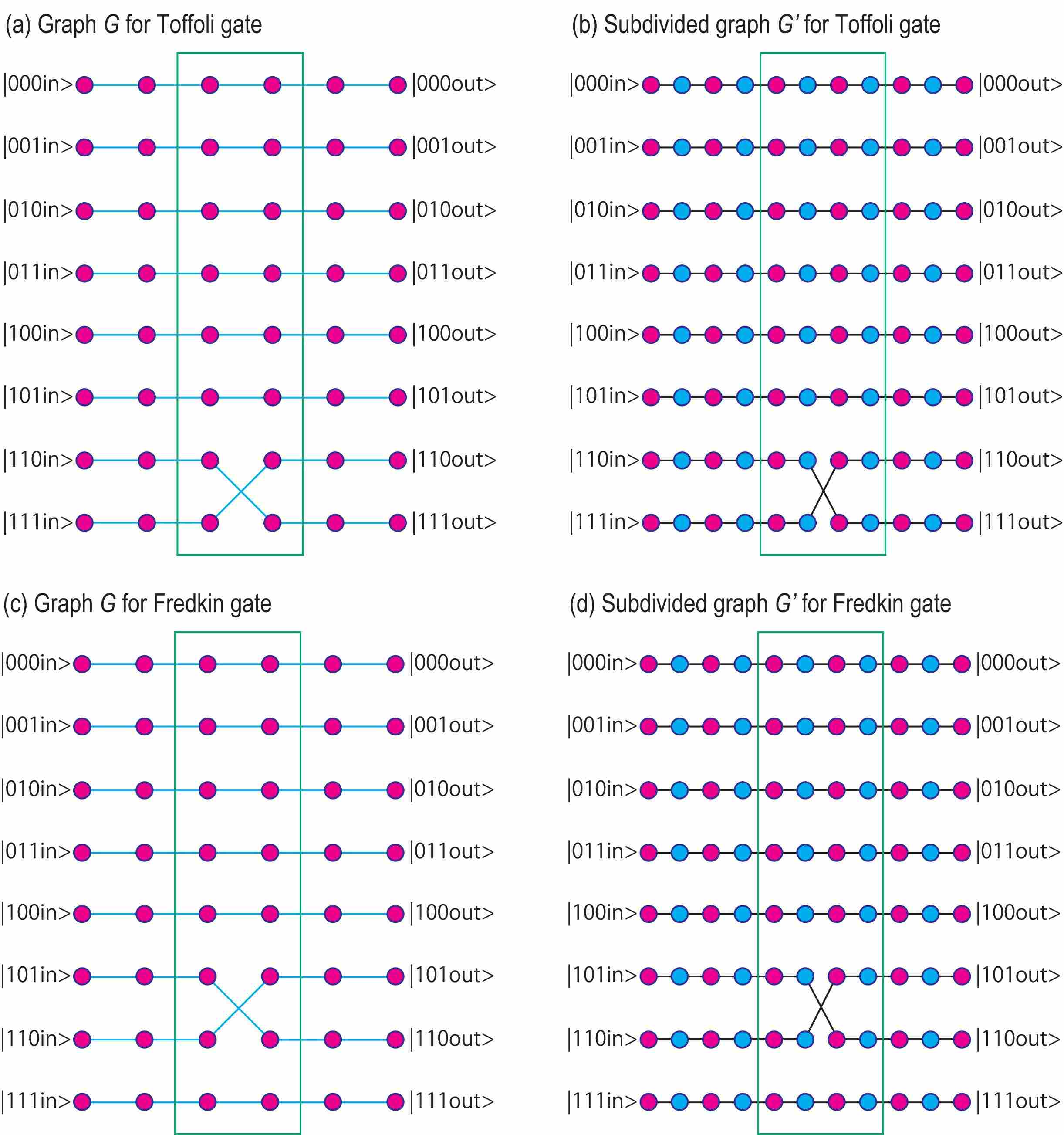}}
\caption{Graph representations of the (a) Toffoli and (b) Fredkin gates.}
\label{FigToffoli}
\end{figure}

\textit{Toffoli gate:} The Toffoli gate is a controlled-controlled NOT gate
defined by%
\begin{equation}
U_{\text{Toffoli}}\equiv \left( 
\begin{array}{cc}
I_{4} & O_{4} \\ 
O_{4} & U_{\text{CNOT}}%
\end{array}%
\right) =\left( 
\begin{array}{cccccccc}
1 & 0 & 0 & 0 & 0 & 0 & 0 & 0 \\ 
0 & 1 & 0 & 0 & 0 & 0 & 0 & 0 \\ 
0 & 0 & 1 & 0 & 0 & 0 & 0 & 0 \\ 
0 & 0 & 0 & 1 & 0 & 0 & 0 & 0 \\ 
0 & 0 & 0 & 0 & 1 & 0 & 0 & 0 \\ 
0 & 0 & 0 & 0 & 0 & 1 & 0 & 0 \\ 
0 & 0 & 0 & 0 & 0 & 0 & 0 & 1 \\ 
0 & 0 & 0 & 0 & 0 & 0 & 1 & 0%
\end{array}%
\right) ,
\end{equation}%
where $I_{4}$ is the four-dimensional identity matrix and $O_{4}$ is the
four-dimensional null matrix with the three-qubit basis $\left\{ \left\vert
000\right\rangle ,\left\vert 001\right\rangle ,\left\vert 010\right\rangle
,\left\vert 011\right\rangle ,\left\vert 100\right\rangle ,\left\vert
101\right\rangle ,\left\vert 110\right\rangle ,\left\vert 111\right\rangle
\right\} ^{t}$. 
\begin{equation}
\left( 
\begin{array}{c}
\left\vert 000\right\rangle _{\text{out}} \\ 
\left\vert 001\right\rangle _{\text{out}} \\ 
\left\vert 010\right\rangle _{\text{out}} \\ 
\left\vert 011\right\rangle _{\text{out}} \\ 
\left\vert 100\right\rangle _{\text{out}} \\ 
\left\vert 101\right\rangle _{\text{out}} \\ 
\left\vert 110\right\rangle _{\text{out}} \\ 
\left\vert 111\right\rangle _{\text{out}}%
\end{array}%
\right) =U_{\text{Toffoli}}\left( 
\begin{array}{c}
\left\vert 000\right\rangle _{\text{in}} \\ 
\left\vert 001\right\rangle _{\text{in}} \\ 
\left\vert 010\right\rangle _{\text{in}} \\ 
\left\vert 011\right\rangle _{\text{in}} \\ 
\left\vert 100\right\rangle _{\text{in}} \\ 
\left\vert 101\right\rangle _{\text{in}} \\ 
\left\vert 110\right\rangle _{\text{in}} \\ 
\left\vert 111\right\rangle _{\text{in}}%
\end{array}%
\right) .
\end{equation}%
We use eight wires for three-qubit circuits. The two wires $\left\vert
110\right\rangle _{\text{in}}$ and $\left\vert 111\right\rangle _{\text{in}}$
are interchanged within the widget, while the others are directly connected
to the outputs as shown in Fig.\ref{FigToffoli}(a).

\textit{Fredkin gate:} The Fredkin gate is a controlled SWAP gate defined by%
\begin{equation}
U_{\text{Fredkin}}\equiv \left( 
\begin{array}{cc}
I_{4} & O_{4} \\ 
O_{4} & U_{\text{SWAP}}%
\end{array}%
\right) =\left( 
\begin{array}{cccccccc}
1 & 0 & 0 & 0 & 0 & 0 & 0 & 0 \\ 
0 & 1 & 0 & 0 & 0 & 0 & 0 & 0 \\ 
0 & 0 & 1 & 0 & 0 & 0 & 0 & 0 \\ 
0 & 0 & 0 & 1 & 0 & 0 & 0 & 0 \\ 
0 & 0 & 0 & 0 & 1 & 0 & 0 & 0 \\ 
0 & 0 & 0 & 0 & 0 & 0 & 1 & 0 \\ 
0 & 0 & 0 & 0 & 0 & 1 & 0 & 0 \\ 
0 & 0 & 0 & 0 & 0 & 0 & 0 & 1%
\end{array}%
\right) .
\end{equation}%
\nobreak Similarly to the Toffoli gate, the two wires $\left\vert
101\right\rangle _{\text{in}}$ and $\left\vert 110\right\rangle _{\text{in}}$
are interchanged within the widget, while the others are directly connected
to the outputs as shown in Fig.\ref{FigToffoli}(b).

\textit{The }$\pi /4$\textit{\ phase shift gate:} In the case of $N=2$, the
momentum is $k_{0}=11\pi /8$. By inserting $m$ inductors in one wire, we may
generate the phase shift $2mk_{0}$ as follows.%
\begin{equation}
\arraycolsep=4mm\renewcommand{\arraystretch}{1.6}%
\begin{tabular}{|c|c|c|c|c|c|c|c|c|}
\hline
$m$ & $1$ & $2$ & $3$ & $4$ & $5$ & $6$ & $7$ & $8$ \\ \hline
$\text{phase shift}$ & $\frac{3\pi }{4}$ & $\frac{3\pi }{2}$ & $\frac{\pi }{4%
}$ & $\pi $ & $\frac{7\pi }{4}$ & $\frac{\pi }{2}$ & $\frac{5\pi }{4}$ & $0$
\\ \hline
\end{tabular}
\label{table}
\end{equation}%
We have chosen $m=3$ in order to realize the $\pi /4$ phase-shift gate in
the main text.

\textit{The }$\pi /8$\textit{\ phase shift gate:} In the case of $N=4$, the
momentum is $k_0=21\pi /16$. By inserting $m$ inductors in one wire, we may
generate the phase shift $2mk_0$ as follows.

\begin{equation}
\arraycolsep=4mm\renewcommand{\arraystretch}{1.6}%
\begin{tabular}{|c|c|c|c|c|c|c|c|c|c|c|c|c|c|c|c|c|}
\hline
$m$ & $1$ & $2$ & $3$ & $4$ & $5$ & $6$ & $7$ & $8$ & $9$ & $10$ & $11$ & $%
12 $ & $13$ & $14$ & $15$ & $16$ \\ \hline
$\text{phase shift}$ & $\frac{5\pi }{8}$ & $\frac{5\pi }{4}$ & $\frac{15\pi 
}{8}$ & $\frac{\pi }{2}$ & $\frac{9\pi }{8}$ & $\frac{7\pi }{4}$ & $\frac{%
3\pi }{8}$ & $\pi $ & $\frac{13\pi }{8}$ & $\frac{\pi }{4}$ & $\frac{7\pi }{8%
}$ & $\frac{3\pi }{2}$ & $\frac{\pi }{8}$ & $\frac{3\pi }{4}$ & $\frac{11\pi 
}{8}$ & 0 \\ \hline
\end{tabular}%
\end{equation}%
We choose $m=13$ in order to realize the $\pi /8$ phase shift gate.

\begin{figure}[t]
\centerline{\includegraphics[width=0.68\textwidth]{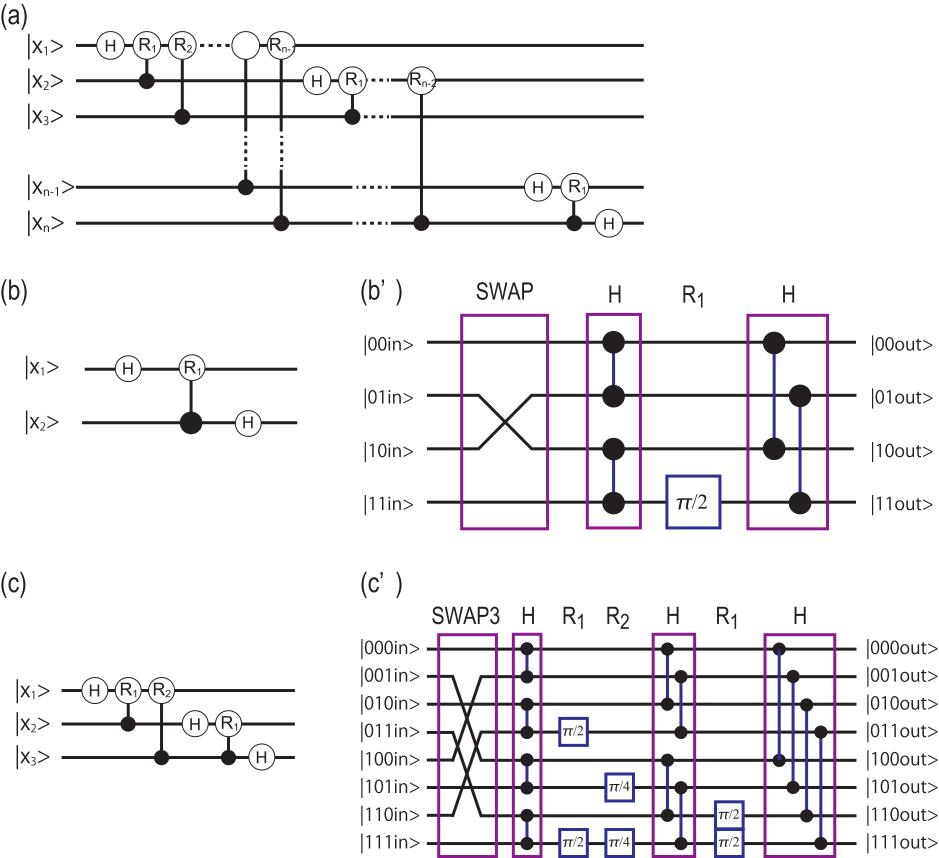}}
\caption{(a) Quantum circuit representation of quantum Fourier
transformation. The symbol $R_{n}$ indicates the $\protect\pi /2^{n}$
phase-shift gate and $H $ indicates the Hadamard gate. (b) Two-qubit quantum
Fourier transformation. (c) Three-qubit quantum Fourier transformation.
(b'), (c') The corresponding reduced graph representation.}
\end{figure}

\textit{QFT with }$N=4$\textit{:} The QFT for $N=4$ is explicitly given by%
\begin{equation}
U_{\text{QFT2}}=\frac{1}{2}\left( 
\begin{array}{cccc}
1 & 1 & 1 & 1 \\ 
1 & i & -1 & -i \\ 
1 & -1 & 1 & -1 \\ 
1 & -i & -1 & i%
\end{array}%
\right) .
\end{equation}

It is decomposed into the sequential applications of the SWAP gate, the
Hadamard gate and the controlled-phase-shift gate $U_{2\rightarrow \phi }$
as 
\begin{equation}
U_{\text{QFT2}}=\left( U_{\text{H}}\otimes I_{2}\right) \left(
U_{2\rightarrow \pi /2}\right) \left( I_{2}\otimes U_{\text{H}}\right) U_{%
\text{SWAP}},
\end{equation}%
where%
\begin{equation}
I_{2}\otimes U_{\text{H}}=\frac{1}{\sqrt{2}}\left( 
\begin{array}{cccc}
1 & 1 & 0 & 0 \\ 
1 & -1 & 0 & 0 \\ 
0 & 0 & 1 & 1 \\ 
0 & 0 & 1 & -1%
\end{array}%
\right) ,\quad U_{\text{H}}\otimes I_{2}=\frac{1}{\sqrt{2}}\left( 
\begin{array}{cccc}
1 & 0 & 1 & 0 \\ 
0 & 1 & 0 & 1 \\ 
1 & 0 & -1 & 0 \\ 
0 & 1 & 0 & -1%
\end{array}%
\right) ,\quad U_{2\rightarrow \pi /2}=\left( 
\begin{array}{cccc}
1 & 0 & 0 & 0 \\ 
0 & 1 & 0 & 0 \\ 
0 & 0 & 1 & 0 \\ 
0 & 0 & 0 & i%
\end{array}%
\right) .
\end{equation}

\textit{QFT with }$N=8$\textit{:} The QFT for $N=8$ is explicitly given by%
\begin{equation}
U_{\text{QFT3}}=\frac{1}{2\sqrt{2}}\left( 
\begin{array}{cccccccc}
1 & 1 & 1 & 1 & 1 & 1 & 1 & 1 \\ 
1 & \omega & \omega ^{2} & \omega ^{3} & \omega ^{4} & \omega ^{5} & \omega
^{6} & \omega ^{7} \\ 
1 & \omega ^{2} & \omega ^{4} & \omega ^{6} & 1 & \omega ^{2} & \omega ^{4}
& \omega ^{6} \\ 
1 & \omega ^{3} & \omega ^{6} & \omega & \omega ^{4} & \omega ^{7} & \omega
^{2} & \omega ^{5} \\ 
1 & \omega ^{4} & 1 & \omega ^{4} & 1 & \omega ^{4} & 1 & \omega ^{4} \\ 
1 & \omega ^{5} & \omega ^{2} & \omega ^{7} & \omega ^{4} & \omega & \omega
^{6} & \omega ^{3} \\ 
1 & \omega ^{6} & \omega ^{4} & \omega ^{2} & 1 & \omega ^{6} & \omega ^{4}
& \omega ^{2} \\ 
1 & \omega ^{7} & \omega ^{6} & \omega ^{5} & \omega ^{4} & \omega ^{3} & 
\omega ^{2} & \omega%
\end{array}%
\right)
\end{equation}%
with $\omega =e^{i\pi /4}$.

It is decomposed into the sequential applications of the three-qubit SWAP
gate $U_{\text{SWAP3}}$, the Hadamard gate and the controlled-phase-shift
gate $U_{n\rightarrow \phi }$ as 
\begin{equation}
U_{\text{QFT3}}=\left( U_{\text{H}}\otimes I_{2}\otimes I_{2}\right) \left(
U_{3\rightarrow \pi /2}\right) \left( I_{2}\otimes U_{\text{H}}\otimes
I_{2}\right) \left( U_{3\rightarrow \pi /4}\right) \left( U_{2\rightarrow
\pi /2}\right) \left( I_{2}\otimes I_{2}\otimes U_{\text{H}}\right) U_{\text{%
SWAP3}},
\end{equation}%
with%
\begin{eqnarray}
U_{2\rightarrow \pi /2} &=&\text{diag.}\left[ 1,1,1,i,1,1,1,i\right] \\
U_{3\rightarrow \pi /4} &=&\text{diag.}\left[ 1,1,1,1,1,e^{i\pi
/4},1,e^{i\pi /4}\right] \\
U_{3\rightarrow \pi /2} &=&\text{diag.}\left[ 1,1,1,1,1,1,i,i\right]
\end{eqnarray}%
and 
\begin{equation}
U_{\text{SWAP3}}\left\vert j_{3}j_{2}j_{1}\right\rangle =\left\vert
j_{1}j_{2}j_{3}\right\rangle
\end{equation}%
with%
\begin{equation}
U_{\text{SWAP3}}=\left( 
\begin{array}{cccccccc}
1 & 0 & 0 & 0 & 0 & 0 & 0 & 0 \\ 
0 & 0 & 0 & 0 & 1 & 0 & 0 & 0 \\ 
0 & 0 & 1 & 0 & 0 & 0 & 0 & 0 \\ 
0 & 0 & 0 & 0 & 0 & 0 & 1 & 0 \\ 
0 & 1 & 0 & 0 & 0 & 0 & 0 & 0 \\ 
0 & 0 & 0 & 0 & 0 & 1 & 0 & 0 \\ 
0 & 0 & 0 & 1 & 0 & 0 & 0 & 0 \\ 
0 & 0 & 0 & 0 & 0 & 0 & 0 & 1%
\end{array}%
\right) .
\end{equation}

\end{document}